\documentclass[sigconf,authorversion,nonacm,screen,natbib=false]{acmart}
\AtBeginDocument{%
  }

\RequirePackage[
  datamodel=acmdatamodel,
  style=acmnumeric,
  ]{biblatex}

\begin{document}

\title{Co-constructing Explanations for AI Systems using Provenance}


\author{Jan-Christoph Kalo}
\affiliation{%
  \institution{University of Amsterdam}
  \city{Amsterdam}
  \country{Netherlands}
}
\email{j.c.kalo@uva.nl}
\author{Fina Polat}
\affiliation{%
  \institution{University of Amsterdam}
  \city{Amsterdam}
  \country{Netherlands}
}
\email{f.yilmazpolat@uva.nl}
\author{Shubha Guha}
\affiliation{%
  \institution{University of Amsterdam}
  \city{Amsterdam}
  \country{Netherlands}
}
\email{s.guha@uva.nl} 
\author{Paul Groth}
\affiliation{%
  \institution{University of Amsterdam}
  \city{Amsterdam}
  \country{Netherlands}
}
\email{p.t.groth@uva.nl}

\renewcommand{\shortauthors}{Trovato et al.}

\begin{abstract}
Modern AI systems are complex workflows containing multiple components and data sources. Data provenance provides the ability to interrogate and potentially explain the outputs of these systems. However, provenance is often too detailed and not contextualized for the user trying to understand the AI system. In this work, we present our vision for an interactive agent that works together with the user to co-construct an explanation that is simultaneously useful to the user as well as grounded in data provenance. To illustrate this vision, we present: 1) an initial prototype of such an agent; and 2) a scalable evaluation framework based on user simulations and a large language model as a judge approach. 
\end{abstract}



\keywords{co-construction of explanations, provenance, interactivity, XAI}


\maketitle

\section{Introduction}

Providing explanations for how complex AI and data science systems produce results is increasingly a critical requirement for their deployment and use \cite{balasubramaniam2023transparency}. To address this need, the area of Explainable AI (XAI) \cite{BARREDOARRIETA202082} has emerged. 
More specifically, in contrast to feature-based explanation approaches, {\em trace-based} explanations~\cite{chari2020directions} aim to explain how a system arrives at results. Trace-based explanations are important because they allow multiple stakeholders, such as developers and product owners, to understand but also improve AI systems \cite{mlpractitioner2019}. Trace-based explanation approaches often employ {\em provenance} either as the explanation itself or to inform the creation of explanations for the end user. 

However, emerging literature argues that explanation is not a single output or even an interactive user interface over such an output, instead it is a process in which the explainer and explainee {\em co-construct} an explanation together \cite{rohlfing2020explanation}. By way of analogy, think of how a student and teacher interact together to help find the explanation that the student needs to understand a new concept. 

Hence, in this work, we describe our vision of explanation agents that make use of provenance to interatively co-construct explanations together with users. We motivate this vision with an initial prototype that uses provenance information in combination with large language models (LLMs) to create an environment for the co-construction of explanations for complex machine learning processes. Importantly, we discuss an approach for evaluating such systems that allows for their scalable assessment. 

\section{Related Work}
\paragraph{Explainable AI:} There is a large amount of work on explainable AI which has been comprehensively covered by recent surveys \cite{schwalbe2024comprehensive,seeliger2019semantic,li2020survey,evalxai2021}. In contrast to feature-based explanations, where the goal is to identify which input features are an important influence on a machine learning model's output, our work is oriented toward what are termed {\em trace-based} explanations~\cite{chari2020directions}, which are explanations based on how a system arrived at a particular outcome.  The goal is to incorporate all of the operations of the system including which data was used, how data preparation occurred, and so on. Again, this is important because biases in the whole of the system can have downstream impacts \cite{schelter2020taming,zhao-etal-2017-men}. Thus, a focus on such explanations seems warranted.

\paragraph{Provenance \& Explanations:} For trace-based explanations, provenance plays a critical role. There are a variety of systems that use provenance for explanations and debugging in complex AI systems. \cite{datacentricexplanations2021,schelter2020taming,Grafberger2022,10.1145/3329486.3329489,10.1145/3514221.3517886}. In general, there is a broad literature on the importance of provenance and its use for XAI. Here, we refer the reader to the recent survey by Kale et al. \cite{kale2023provenance}. Our work is orthogonal in that we aim to use provenance as an input to a co-construction process, allowing users to understand how a system has arrived at a conclusion by interactively engaging with a user. The closest work to our approach is that of \cite{Daga2024} which looks as explanation as exploration. However, they did not look at tailoring the explanation to the user, instead focusing on how users can explore an abstract version of data science pipelines, 

\paragraph{Evaluating Explanations:}
A key question is how to evaluate explanations in a co-construction scenario. A variety of approaches with corresponding metrics have emerged in the literature. According to \cite{schwalbe2024comprehensive}, metrics can be broadly categorized as:

\begin{itemize}
\item functionally-grounded - metrics that do require human feedback and measure properties of the explanation (e.g. faithfulness - how accurately does the explanation correspond to the thing being explained);
\item human-grounded metrics - metrics that involve human participation either through feedback, observation or proxy tasks (e.g. how interpretable is an explanation to an end user);
\item application-grounded - metrics that are based on an explanation's usage in an application (e.g. does the performance of the human-AI system improve on a downstream task).
\end{itemize}

Given that we aim for interactivity, approaches for evaluating dialogue systems are also relevant. One natural approach for the evaluation of dialogues is human-based evaluation, which includes lab-based experiments, observations of users in the field, and crowd-sourcing \cite{deriu2021survey}. However, as \cite{deriu2021survey} argues, such evaluations are difficult to carry out and not necessarily repeatable, and human feedback may not be reliable. Hence, the dialogue systems community has investigated automating all or parts of evaluation. Dialogue quality is defined by task success (similar to application-grounded metrics) and efficiency of dialogue \cite{deriu2021survey}. Common evaluation methods employed are user satisfaction modeling - creating models which can rate dialogues in a manner that mirrors human judges - and user simulation \cite{deriu2021survey}.

In our vision, we build on all three lines of work, investigating how to use provenance information to support co-construction of explanations for XAI systems and how to evaluate such systems.

\section{Prototype}
\begin{figure}
    \centering
    \includegraphics[width=\columnwidth]{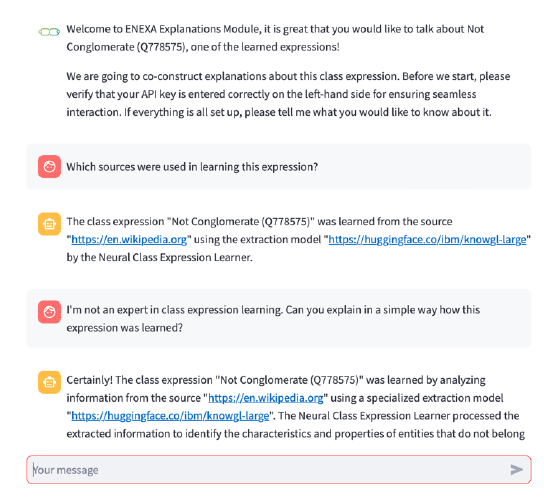}
    \caption{Co-constructing the explanation of the results of class expression learning with the ENEXA explanation chatbot.}
    \label{fig:enexachat}
\end{figure}

We have developed a prototype interactive chatbot (ENEXA explanation chatbot) that aims to co-construct an explanation of the outcomes of a machine learning system. Figure \ref{fig:enexachat} shows the interface for the system.

Specifically, we use the scenario of neural class expression learning \cite{neuralClassExpressionLearning2023}. This is the task of learning a new description of an ontology class given positive and negative examples of instances of that class. In this scenario, the provided examples are sourced from an information extraction pipeline. In the prototype, class expressions are verbalised and presented to the user. The user can directly ask questions about the class expressions.

\begin{figure}
    \centering
    \includegraphics[width=\columnwidth]{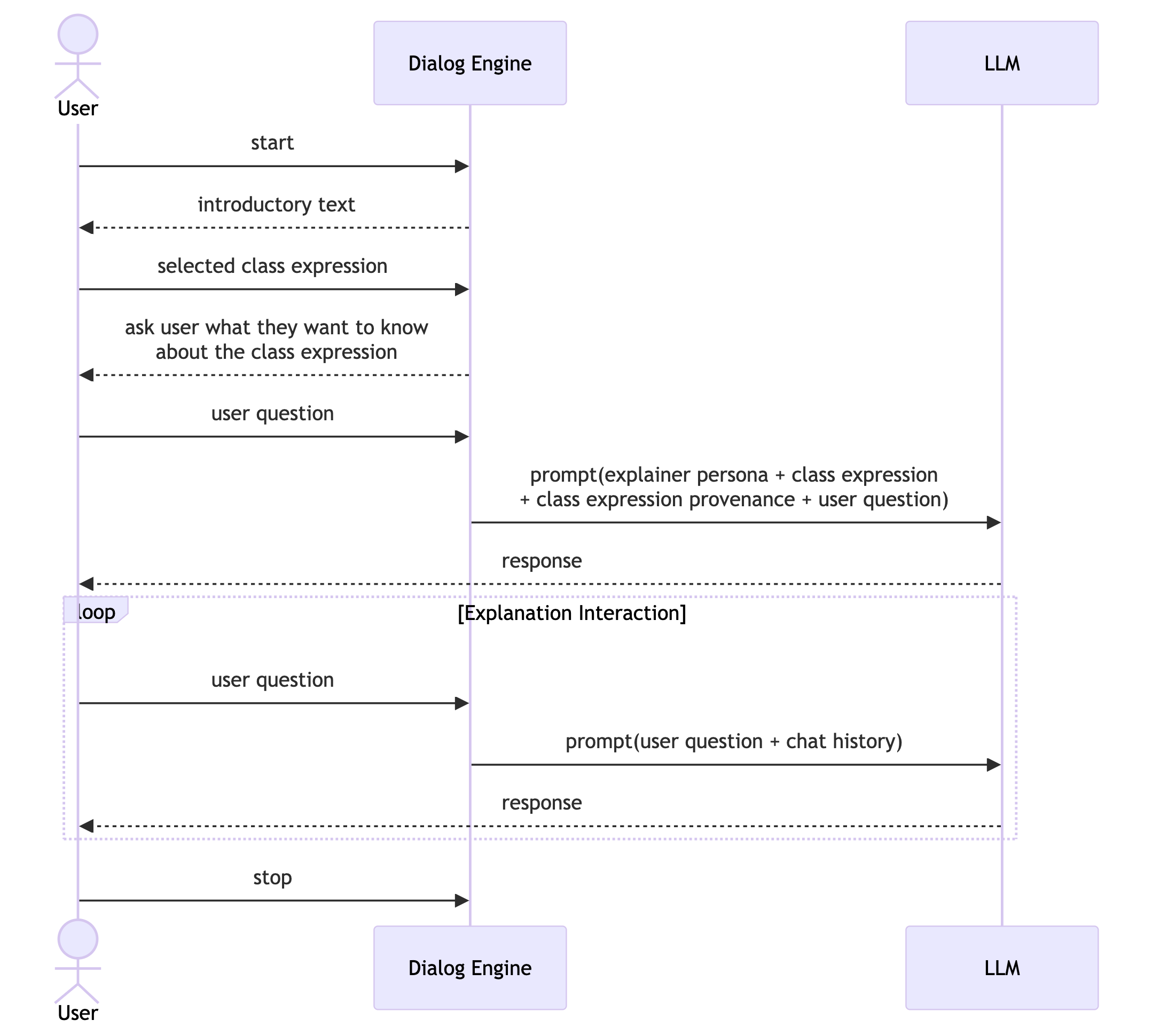}
    \caption{Sequence diagram for the ENEXA chatbot used for co-construction of explanations for a neural class expression learner.}
    \label{fig:chatseq}
\end{figure}

Figure \ref{fig:chatseq} provides a sequence diagram that illustrates the architecture of the system. 

The Dialog Engine mediates between the user and an underlying large language model (LLM). In particular, it maintains the chat history but also provides additional information about the provenance of how the particular class expression was learned. Specifically, it provides provenance indicating the underlying data sources as well as the information extraction procedure used. This provenance information is added to the prompts as input to the LLM. The prompts employed are available in the source code repository.\footnote{Source code and Docker files are available at: \url{https://github.com/FinaPolat/XCE-Streamlit-V3/}}

The prototype illustrates how the explanation can be tailored to the user and how specific parts of the provenance can be focused on (e.g. the sources). Importantly, the provenance can be mixed with relevant contextual knowledge from the large language model (e.g. more or less detail in explaining how class expression learning works). 

\section{Evaluation Framework}
A critical challenge is how to evaluate these systems. Given the fact that we would like to have an explanation system that tailors explanations towards a wide variety of users and a potentially large number of machine learning tasks, end-user evaluation is not feasible. Following the related work, an automated evaluation technique seems to be an effective means forward. 

To tackle this challenge, we turn towards the recent LLM-as-a-judge approach \cite{llmjudgeNeurips2023}.  In multi-turn dialogues - which is our use case - \cite{llmjudgeNeurips2023} showed high levels of agreement between LLM agents and human experts. Beyond acting as judge of dialogue quality, LLMs can also be used within dialogue systems to simulate users \cite{braggaar2024evaluatingtaskorienteddialoguesystems}. This builds on the emerging work in using LLMs to simulate personas to help with tasks such as annotation \cite{hu-collier-2024-quantifying}, social science experiments \cite{horton2023large} and survey responses \cite{moon-etal-2024-virtual}. It should be noted that in using these techniques there are concerns as to whether the correlations to human performance seen in studies can be seen as truly viable proxies for human performance \cite{tjuatja-etal-2024-llms}. For a detailed review of these critiques in the context of dialogue evaluation as well as a comprehensive survey, we refer the reader to \cite{braggaar2024evaluatingtaskorienteddialoguesystems}. Based on the requirements to evaluate multiple personas and multiple explanation results, we believe a user simulation approach in combination with an LLM-as-a-judge is a promising path forward.

To investigate this idea, we developed an automated chatbot explanation and evaluation framework.\footnote{Source code is available at: \url{https://github.com/JanKalo/enexa_explanation/}} The framework is able to simulate different user personas using LLMs and have them interact with an explainer agent as described above. The user persona agents are prompted LLMs with different engagement styles. Subsequently, the trace of the explanation is then evaluated by a separate agent based on a series of input criteria.  Currently, we employ the following criteria using a 5 point scale.

\begin{enumerate}    
\item \textbf{Clarity \& Structure}: Does the explanation flow logically? Is it easy to follow?
\item \textbf{Depth \& Completeness}:  Does the explanation offer sufficient detail without omitting crucial points?
 \item \textbf{Correctness \& Fidelity}: Are facts accurate, and does the explanation remain faithful to the original query/context?
\item \textbf{Relevance \& Focus}: Does the content stay on-topic and address user queries directly?
\item \textbf{Appropriateness for the Persona}: Is the style/tone appropriate for the user's persona (e.g., an AI engineer, business strategist, etc.)?
\item \textbf{Transparency}: Does the explanation clarify its reasoning or highlight uncertainties?
\item \textbf{Engagement \& Intuition}: Is the conversation engaging, and does it address the user’s interests intuitively?
\end{enumerate}

The framework maintains the independence of each agent, and the interaction between agent and explainer goes through a series of rounds. The total number of interactions between the agent and the explainer is limited via a parameter. The framework works with both closed models (e.g. GPT-4) as well as open models (Mistral). The framework allows us to scale up evaluation and analysis of different explanation approaches, to discern potential differences between approaches, and to try different criteria.

\section{Experiments}
Following from the prototype, we experimented with 6 different chat personas using 10 different class expressions with associated metadata. Note that the personas are meant to broadly indicate a role and not necessarily substitute for true human behaviour. We limit the interaction to three turns. An example of a trace that is being evaluated is shown in Figure \ref{fig:explantiontrace}.

\begin{figure}
\scriptsize
\tt
\begin{itemize}
    \item \textbf{User:} What criteria were used to define 'Oscar-winning Actor (Q10800557)'?
    \item \textbf{Assistant:} The criteria used to define 'Oscar-winning Actor (Q10800557)' were based on identifying actors who have won an Oscar. Positive examples, such as Leonardo DiCaprio and Meryl Streep, are actors noted for having received this accolade. Negative examples, like Keanu Reeves and Harrison Ford, are actors who have not won an Oscar. This classification was derived using the Neural Class Expression Learner model, and the information was verified from Wikipedia.
    \item \textbf{User:} What if an actor has been nominated but hasn't won an Oscar? How is that handled?
    \item \textbf{Assistant:} If an actor has been nominated but hasn't won an Oscar, they would not be classified under 'Oscar-winning Actor (Q10800557)'. The criteria specifically focus on actors who have won the Oscar, as demonstrated by negative examples like Keanu Reeves and Harrison Ford, who are actors with notable careers but have not won the award. Therefore, only winning the Oscar qualifies an actor for this classification.
    \item \textbf{User:} What about an actor who has won an Oscar for a non-acting category? How does the class expression handle this?
    \item \textbf{Assistant:} Based on the provided metadata, the class expression 'Oscar-winning Actor (Q10800557)' specifically focuses on actors who have won an Oscar in acting roles. The positive examples, Leonardo DiCaprio and Meryl Streep, are recognized for their acting achievements. The metadata does not provide information on how it handles actors who have won Oscars in non-acting categories, suggesting the primary focus is on acting categories only.
\end{itemize}
\caption{Example of a chat evaluated with the persona `data skeptic.' \label{fig:explantiontrace}}
\end{figure}

The results of the experiments are shown in Table \ref{tab:chat_persona_metrics}. We see in general that the explainer is able to remain faithful to the input metadata about the class expression (i.e. correctness \& fidelity). The explainer is also able to stay on-topic (i.e. relevance \& focus). However, there is room for improvement in terms of engagement. We think depth could be improved by a longer number of turns. In terms of personas, scores appear to be better for the technical personas (e.g. data skeptic and AI engineer). This could be because of the highly technical nature of class expression learning.

\begin{table*}
    \centering
    \begin{tabular}{p{1.5cm}ccccccc}
        \hline
        Chat  & Clarity & Depth & Correctness &  Relevances & Appropriateness & Transparency & Engagement \\
         Persona & \& Structure &  \& Completeness &  \& Fidelity &  \& Focus & for the Persona &  &  \& Intuition \\
        & Mean (Std) & Mean (Std) & Mean (Std) & Mean (Std) & Mean (Std) & Mean (Std) & Mean (Std) \\
        \hline
        ai engineer & 4.2 (0.84) & 3.8 (1.30) & 4.8 (0.45) & 4.8 (0.45) & 4.2 (0.84) & 3.8 (0.45) & 3.8 (1.30) \\
        business strategist & 3.8 (1.10) & 3.0 (1.41) & 4.2 (1.10) & 3.8 (1.79) & 3.8 (1.30) & 3.8 (0.45) & 3.2 (1.64) \\
        curious citizen & 3.8 (0.45) & 3.8 (0.45) & 5.0 (0.00) & 4.6 (0.55) & 4.0 (0.00) & 3.4 (0.89) & 3.4 (0.55) \\
        data skeptic & 4.4 (0.55) & 4.0 (0.71) & 5.0 (0.00) & 4.8 (0.45) & 4.2 (0.45) & 4.0 (0.71) & 4.0 (0.71) \\
        domain expert & 4.2 (0.45) & 3.4 (0.55) & 4.8 (0.45) & 4.6 (0.55) & 4.0 (0.00) & 3.6 (0.55) & 3.4 (0.55) \\
        knowledge graph specialist & 4.0 (0.71) & 4.0 (0.71) & 4.8 (0.45) & 4.6 (0.89) & 4.2 (0.45) & 3.8 (0.45) & 3.8 (0.45) \\
        \hline
    \end{tabular}
    \caption{Evaluation results for different chat personas. Scale from 1 to 5 on each dimension, with 5 being the best.  \label{fig:expPersonaEval}}
    \label{tab:chat_persona_metrics}
\end{table*}

\section{Vision and Conclusion}
Co-construction of explanations is an emerging way to think about explaining complex AI systems. Provenance provides a foundation for such systems by detailing how the outputs of such systems are produced and what data they use and create. Our initial prototype illustrates the ability to tailor an explanation to an end-user based on provenance information in a back-and-forth dialogue. This allows the explanation chatbot to incrementally focus on relevant pieces of provenance while also incorporating the background knowledge of the underlying LLM through the interaction with the end-user. 

A critical question is how to evaluate these systems, given that they need to support many user personas and many potential systems and tasks. Here we propose the combination of user simulation with an LLM-as-a-judge to provide a framework for large-scale reproducible evaluations. 

Going forward, in terms of evaluation, we envision the need for a battery of hundreds to thousands of well-described personas that can be used for evaluation. Furthermore, an expansion of the number of criteria that can be employed. Additionally, we believe that seeing if the assessments produced by the evaluation framework correlate to application-grounded metrics is a promising direction. 

With respect to co-constructing explanations based on provenance, we envision taking advantage of much more detailed provenance information as well as the ability to use multi-modal outputs to explain the traces of a system to the user. Imagine, for example, an agent, $m$, that can illustrate and gesture to particular diagrammatic parts of a system on the fly while walking the user through how data traversed the various components of the AI system. 

From a provenance systems perspective, there are many avenues for future work. For example, implementing retrieval augmented generation over a provenance store to effectively select relevant provenance for the explanation. Another opportunity is to integrate LLMs directly in the collection and preparation of the necessary provenance. \cite{Lauro2024}. Additionally, developing algorithms to provide the appropriate level of abstraction for the explanation agent to best interact with the user is of interest, especially as many provenance systems and representation languages provide this capability \cite{herschel2017survey,moreau2022provenance}.

Our vision is of interactive environments where users interact with agents that can explain how AI systems produced their results on their terms, constrained and underpinned by rich provenance.

\begin{acks}
This work was partially supported by the EU’s Horizon Europe research and innovation programme within the ENEXA project (grant Agreement no. 101070305).
\end{acks}

\printbibliography

\appendix

\end{document}